\documentstyle[12pt,epsf]{article} 
\begin{document}
\baselineskip 16pt plus 2pt minus 2pt
\newcommand{\beq}{\begin{equation}}
\newcommand{\eeq}{\end{equation}}
\newcommand{\beqa}{\begin{eqnarray}}
\newcommand{\eeqa}{\end{eqnarray}}
\newcommand{\dfrac}{\displaystyle \frac}
\renewcommand{\thefootnote}{\#\arabic{footnote}}
\newcommand{\ve}{\varepsilon}
\newcommand{\krig}[1]{\stackrel{\circ}{#1}}
\newcommand{\barr}[1]{\not\mathrel #1}
\begin{titlepage}


\hfill KFA--IKP(TH)--1997--01


\begin{center}

\vspace{2.0cm}

{\Large  \bf { Strange vector currents and the OZI--rule}} \\

\vspace{1.2cm}
                              
{\large Ulf-G. Mei\ss ner$^{\ddag,1)}$, V. Mull$^{\ddag}$
  \footnote{Present address: IBM Informationssysteme GmbH, K\"oln, Germany.} , 
J. Speth$^{\ddag,2)}$, J. W. Van Orden$^{\dag,3)}$}

\vspace{0.7cm}

$^{\ddag}$Forschungszentrum J\"ulich, IKP (Theorie), D--52425 J\"ulich, Germany

$^{\dag}$Department of Physics, Old Dominion University, Norfolk, VA
23529, USA\\ and\\
Jefferson Lab., 12000 Jefferson Ave., Newport News, VA23106, USA

\vspace{0.9cm}

{\it
email:\\ $^{1)}$Ulf-G.Meissner@kfa-juelich.de, $^{2)}$J.Speth@kfa-juelich.de,
$^{3)}$vanorden@jlab.org
}

\end{center}

\vspace{1.5cm}

\begin{center}

 ABSTRACT

\end{center}

\vspace{0.1cm}

\noindent We investigate the role of correlated $\pi\rho$ exchange
in the extraction of matrix elements of the strange vector current in
the proton. We show that a realistic isoscalar spectral function including
this effect leads to sizeably reduced strange vector form factors based
on the dispersion--theoretical analysis of the nucleons' electromagnetic
form factors.

\vspace{2.5cm}

\centerline{DEDICATED TO THE MEMORY OF KARL HOLINDE}

\vfill

\end{titlepage}

One of the outstanding problems in the understanding of the nucleon structure
concerns the strength of various strange operators in the proton. A dedicated
program at Jefferson Laboratory supplemented by experiments at BATES (MIT)
and MAMI (Mainz) is aimed at measuring the form factors related to the strange
vector current $\bar s \gamma_\mu s$ in the nucleon. 
It was already pointed out a long time ago
by Genz and H\"ohler \cite{geho} that the dispersion--theoretical analysis
of the nucleons' electromagnetic form factors allows one to get bounds on
the violation of the OZI rule, which leads one to expect that strange
matrix elements should be small \cite{ozi}. 
This rule has, however, never firmly been
routed in QCD but can be understood qualitatively in large 
N$_C$ (with N$_C$ the number of colors) \cite{witten}. 
Jaffe \cite{bob} showed that under certain assumptions 
the information encoded in the isoscalar nucleon form factors can be used
to extract strange matrix elements. Of particular importance for this type
of analysis is the identification of the two lowest poles in the isoscalar 
spectral function with the $\omega (782)$ and the $\phi (1020)$ mesons. The
corresponding strange form factors turn out to be rather large in magnitude,
related to the strong coupling of the $\phi$ to the nucleon found in the 
dispersion--theoretical analysis \cite{hoeh76}. This analysis was later
updated and extended in \cite{hmd} based on the novel form factor fits
presented in \cite{mmd}. Loosely spoken, such an analysis is based on
a ``maximal'' violation of the OZI rule because the spectral function in the
mass region of about 1~GeV is assumed to be given entirely by the $\phi$ pole.
On the other hand, the coupling of various  mesons (like the $\omega$
and the $\pi$) to the nucleons has been investigated in great detail in the
framework of the Bonn--J\"ulich meson exchange potential for the 
nucleon--nucleon interaction by Holinde and coworkers \cite{ho1} . 
In particular, the correlated $\pi \rho$ \cite{ho1}\cite{jans} and $\pi\pi$ exchange
\cite{kim} has recently been included  consistently. This leads to a 
more realistic microscopic
picture of the isoscalar spectral function in the mass region of the $\phi$
\cite{mull}. Our aim is to combine this novel results from the $NN$ 
interaction with a fit to the nucleon electromagnetic form factors and 
to elucidate the strength of the $\phi NN$ couplings, i.e. the violation 
of the OZI rule, and the consequences for the extraction of the strange 
form factors.

To be specific, consider first the nucleon--nucleon interaction. Although
QCD is believed to be the theory underlying the strong interactions, in the
non--perturbative regime of low-- and medium-energy physics, mesons and baryons
have retained their importance as effective, collective degrees of freedom
for a wide range of nuclear phenomena. This is most apparent for the $NN$ 
system. Here, the interaction between the two nucleons is generated by 
meson exchange \cite{broja}. Resulting potentials, e.g. the 
Paris~\cite{paris}, Nijmegen~\cite{nij} or Bonn~\cite{bonn} potentials, are
able to describe the $NN$ data below pion threshold in a truly quantitative 
manner. The full Bonn potential contains apart from single--meson exchanges
higher--order diagrams involving also the $\Delta$--isobar. The strength of
the various baryon--meson vertices is parametrized by coupling constants.
In addition, form factors with cut--off masses $\Lambda_\alpha$ are included
as additional parameters, they take into account the corresponding vertex 
extensions. However, there are some longstanding conceptual problems hidden
in the choice of parameters which have been resolved in the past years by
Holinde and coworkers. First, the fictitious scalar--isoscalar meson 
$\sigma_{\rm OBE}$, which is needed to provide the intermediate--range
attraction, has been replaced by correlated $2\pi$--exchange \cite{kim}. A
second longstanding discrepancy existed for the cut-off $\Lambda_{\pi NN}$,
which is rather large in the present--day potential models ($\sim 1.3\,$GeV)
compared to the information from other sources, like e.g. $\pi N$ scattering.
In \cite{ho1} it has been shown that the interaction between a $\pi$ and a
$\rho$ meson (correlated $\pi \rho$ exchange) has a strong influence on the
$NN$--potential in the pion channel. It provides a sizeable contribution
with a peak around $1.1\,$GeV. Due to this additional $\pi$--like contribution
one is able to reduce the cut-off $\Lambda_{\pi NN}$, which is now in much
better agreement with information from other sources. The third well--known
discrepancy is the $\omega NN$ coupling constant, which in most of the
$NN$ potentials is three times bigger than predicted by SU(3) symmetry 
\cite{omff}. It has also been shown in \cite{jans} that in the
$\omega$--channel the correlated $\pi \rho$ exchange gives a sizeable
contribution which allows the choice of a value for $g_{\omega NN}$ which is in
reasonable agreement with the SU(3) prediction. The process missing in the
original Bonn potential is depicted in fig.~1. It has been analyzed in
detail in \cite{jans}. The result is given in terms of a dispersion integral,
which for simplicity can be represented by an effective one--boson exchange,
denoted as $\omega '$, in the $\omega$--channel,
\begin{equation} \label{omnnp}
{1\over \pi } \int_{(M_\pi+M_\rho )^2}^\infty {\rho_P^\omega (s,t') \over
t' - t} \, dt' = -{g_{\omega ' NN}^2 \over t - M_{\omega '}^2} \,\,\,  .
\end{equation}
The spectral function $\rho_P^\omega$ is again peaked around $1.1\,$GeV.
The pole fit, eq.(\ref{omnnp}), gives the following  $\omega '$ parameters:
$M_{\omega '} = 1.12\,$GeV, $g_{\omega ' NN}^2/4\pi = 8.5$ for the vector
and $f_{\omega ' NN}^2/4\pi = 1.5$ for the tensor coupling. Moreover, it turns
out that $g_{\omega ' NN} < 0$ and $\kappa_{\omega '} = f_{\omega ' NN} /
g_{\omega ' NN} > 0$. In what follows, we use this 
effective pole instead of the full spectral function.  This approach has 
further been extended to include kaon loops and
hyperon excitations, with the parameters fixed from a study of the
reactions $p \bar p \to \Lambda \bar \Lambda$ and $p \bar p \to 
\Sigma \bar \Sigma$ \cite{mull2}. 
There are sizeable cancellations between the various
contributions from graphs (see fig.~2) with
intermediate $K$'s, $K^*$'s and diagrams with the direct hyperon
interactions \cite{mull} leading to a very small $\phi$ coupling,
\begin{equation} 
\label{coup2}
\frac{g_{\phi NN}^2}{4\pi} \simeq 0.005 \, , \,\, \, \kappa_\phi
\simeq \pm 0.2 \,\, .
\end{equation}
The various contributions are tabulated in table~1.
The sign of the tensor coupling is very sensitive to the details of
the calculation. The smallness of these couplings amounts to a ``resurrection''
of the OZI rule. Note that such strong cancellations have also been 
observed in the quark model study of \cite{isge}. 

The structure of the nucleon as probed with virtual photons is parametrized
in terms of the so--called Dirac ($F_1$) and Pauli ($F_2$) form factors.
These form factors have been measured over a wide range of space--like 
momentum transfer squared, $t = 0 \ldots -35\,$GeV$^2$
but also in the time--like region either in $\bar p p$ annihilation  or in
$e^+ e^- \to \bar p p, \bar n n$ collisions. The tool to analyze these
data in a largely model--independent fashion is dispersion 
theory \cite{hoeh76}. We therefore briefly review the dispersion--theoretical 
formalism developed in \cite{mmd} and discuss the pertinent modifications due 
to the constraints from the $NN$ interaction described before. Assuming the 
validity of unsubtracted dispersion relations for the four form factors 
$F_{1,2}^{(I=0,1)}(t)$ \cite{nota}, one separates the spectral functions 
of the pertinent form factors into a hadronic (meson pole) and a quark 
(pQCD) component as follows,
\begin{equation}
F_i^{(I)} (t) = \tilde{F}_i^{(I)} (t) L(t) = \Bigg[
\tilde{F}_i^\rho (t) \,\delta_{I1} + \sum_{I}
\frac{a_i^{(I)} \, L^{-1}(M^2_{(I)})}{M^2_{(I)} - t }\Bigg] \, \Bigg[
\ln \Bigg( \frac{\Lambda^2 - t}{Q_0^2} \Bigg)\Bigg]^{-\gamma} 
\label{ffism} \end{equation}
where $F_i^\rho(t) = \tilde{F}_i^\rho (t)\, L(t)^{-1}$ parametrizes the 
isovector ($I=1)$ two--pion contribution (including the one from 
the $\rho$) in terms 
of the pion form factor and the P--wave $\pi \pi \bar N N$partial wave 
amplitudes  in a parameter--free manner. In addition, we
have three isovector poles, the masses of the first two can be identified
with physical ones, i.e. $M_{\rho'} = 1.45\,$GeV and  $M_{\rho''} = 1.65\,$GeV.
In the isoscalar channel ($I=0)$, we have the poles representing the 
$\omega$, the $\phi$, the $\omega '$ (parametrizing the correlated 
$\pi\rho$ exchange)  and a fourth pole (denoted $S$). In what follows, 
we will assume that from
these only the $\phi$ and the $S$ couple to strangeness. Notice that
it has recently been shown that there is no enhancement close to
threshold of the isoscalar spectral function due to pion loops \cite{bkmthr}.
Furthermore, $\Lambda \sim 10\,$GeV$^2$ \cite{mmd} separates the hadronic from the 
quark contributions, $Q_0$ is 
related to $\Lambda_{\rm QCD}$ and $\gamma$ is the anomalous dimension,
\begin{equation}
F_i (t) \to 
(-t)^{-(i+1)} \, \biggl[ \ln\biggl(\frac{-t}{Q_0^2}\biggr)
\biggr]^{-\gamma} \, , \quad \gamma = 2 + \frac{4}{3\beta}
\, \, , \quad i = 1,2 \, \, ,
\label{fasy}
\end{equation}
for $t \to -\infty$ and  $\beta$ is the one loop QCD $\beta$--function.
In fact, the fits performed in \cite{mmd} are rather
insensitive to the explicit form of the asymptotic form of the spectral
functions. To be specific, the additional factor
$L(t)$ in Eq.(\ref{ffism}) contributes to the spectral functions
for $t > \Lambda^2$, i.e. in some sense parametrizes the intermediate
states in the QCD regime, above the region of the vector mesons. The
particular logarithmic form has been chosen for convenience.
Obviously, the asymptotic behaviour  is obtained by
choosing the residues of the vector meson pole terms such that the
leading terms in the $1/t$--expansion cancel. In practice, the
additional logarithmic factor is of minor importance for the fit to
the existing data. The number of isoscalar and isovector poles in 
Eq.(\ref{ffism}) is determined by the stability criterion discussed in 
detail in \cite{hoeh76}\cite{mmd}. In short,
we take the minimum number of poles necessary to fit the data. Specifically,
we have four isoscalar and three isovector poles. This fourth isoscalar 
pole is necessary since most isoscalar couplings are fixed (as described 
above) and otherwise
we would not be able to fulfill the various normalization and superconvergence
relations. We are left with three fit parameters, these are the masses of the
third isovector and the fourth isoscalar pole as well as the residuum
$a_1^{\omega}$. 

The spectral functions of the isoscalar form factors $F_{1,2}^{(0)}$ encode 
information about the strange vector current since the photon couples
to a certain extent via mesons with strangeness (here the $\phi$ and the
$S$) to the nucleon. Assuming that the strange form factors have the same 
large momentum fall--off as the isoscalar electromagnetic ones 
\cite{bob}\cite{hmd} and neglecting the small $\omega-\phi$ mixing, 
it is straightforward to extract the strange Dirac and Pauli 
form factors following the formalism outlined in \cite{bob}\cite{hmd}
\begin{eqnarray}
&&F_1^s (t) = t\,L(t)\,a_1^\phi L_\phi^{-1} \, \frac{M_\phi^2 - M_S^2}{(t
- M_\phi^2)(t-M_S^2)}\, , \nonumber \\
&&F_2^s (t) = L(t)\,a_2^\phi L_\phi^{-1} \, \frac{M_\phi^2 - M_S^2}{(t
- M_\phi^2)(t-M_S^2)}\, , \,\,
\end{eqnarray}
with $ L_\phi^{-1} = 1/L(M_\phi^2)$. Clearly, the size of these
strange form factors is given by the strength
of the $\phi$--nucleon couplings (as encoded in the residua $a_{1,2}^\phi$).
In particular, we notice that the sign of the strange radius $r_{1,s}^2$ is
determined from the sign of $a_1^\phi$ whereas the sign of the strange
magnetic moment, $\mu_s = F_2^{(s)}(0)$, is fixed by the sign of the 
tensor coupling $\sim a_2^\phi$.

A best fit to the available data as compiled in \cite{hmd2} is obtained
with $M_{S'} = 1.63\,$GeV, $M_{\rho'''} = 1.72\,$GeV and $a_1^{\omega}
= 0.677$ (for $g_{\phi NN} = -0.24$ and $\kappa_\phi = 0.2$). 
The $\chi^2$/datum of the fit is 1.02. All contraints are fulfilled
to high numerical accuracy. A detailed account of these results is given in
\cite{msvo}. The corresponding strange form factors are shown in fig.~3. 
Notice that  $F_1^{(s)} (t)$ varies very weakly between $t = -1 \ldots
 -10\,$GeV$^2$. Furthermore, the strange  magnetic moment and radius are 
$\mu_s = 0.003 \,{\rm n.m.}$ and $r_s^2 = 0.002\, {\rm fm}^2$, respectively. 
These are orders of magnitude smaller than in previous analysis 
\cite{bob}\cite{hmd} where the $\phi$ pole subsumed the non--strange physics
of the isoscalar spectral function in the mass region of about 1~GeV, i.e. the
sizeable effect of the $\pi\rho$ correlations. 
The use of a more realistic spectral function based on the constraints 
from the $NN$ interaction indeed leads to a reduction of the strange 
matrix elements as anticipated from the OZI rule. Of course, the analysis
presented here can be sharpened by including the effects of $\omega-\phi$ 
mixing and by studying the dependence on the large-$t$ behaviour of the
strange form factors. This will, however, not change the main conclusion of
our work, namely that the inclusion of correlated $\pi\rho$--exchange in the
isoscalar spectral function for the nucleon electromagnetic form factors
leads to a sizeable reduction of the strange vector current matrix elements
in the proton. The experimental information concerning the strange
form factors is thus eagerly awaited for.    
  
\bigskip\bigskip

\noindent {\Large{\bf Acknowledgements}}

\medskip

\noindent We thank Nathan Isgur for useful comments.

\bigskip\bigskip\bigskip

\noindent {\Large{\bf Figure Captions}}

\smallskip

\begin{enumerate}
\item[Fig.1] Correlated $\pi\rho$ exchange missing in the Bonn potential. 
 
\item[Fig.2] Hadronic model for the $\phi NN$ vertex consisting of Born
terms and interaction diagrams (as indicated by the blobs).

\item[Fig.3] Strange form factors $F_1^{(s)} (t)$ (solid line) 
and $F_2^{(s)} (t)$ (dashed line).

\end{enumerate}

\newpage

\section*{Tables}

\bigskip

\renewcommand{\arraystretch}{1.3}

\begin{center}

\begin{tabular}{|c||c|c|}
\hline
\hline
                           &    g      &    f     \\
\hline
$\bar K K$                 &  $-$0.32    &   $-$0.15  \\
$\bar K^* K - \bar K K^*$  &  $-$0.39    &   $-$0.20  \\
$\bar K^* K^*$             &  +0.25    &   +0.40  \\
\hline
all mesons                 &  $-$0.47    &   +0.14  \\
\hline
$\Sigma \bar\Sigma + \Lambda \bar \Lambda$ & +0.23 & $-$0.19 \\
\hline
Sum                     &    $-$0.24   &   $-$0.05   \\
\hline
\hline
\end{tabular}\bigskip

\end{center}

\noindent Table~1: Various contributions to the $\phi NN$ vector (g)
and tensor coupling (f) (compare fig.~2). Given is also the sum
of the individual contributions. In the case of the tensor coupling,
the meson contributions can not
simply be added since the spectral functions have different signs and
the corresponding monopole fit to the sum of the spectral functions
has a considerably harder form factor than the individual
contributions. In contrast, for the vector coupling the monopole cut
offs are all of comparable size and thus one can essentially add the 
individual contributions.

\vskip 1cm

\section*{Figures}


\hskip 1.6in
\epsfysize=3in
\epsffile{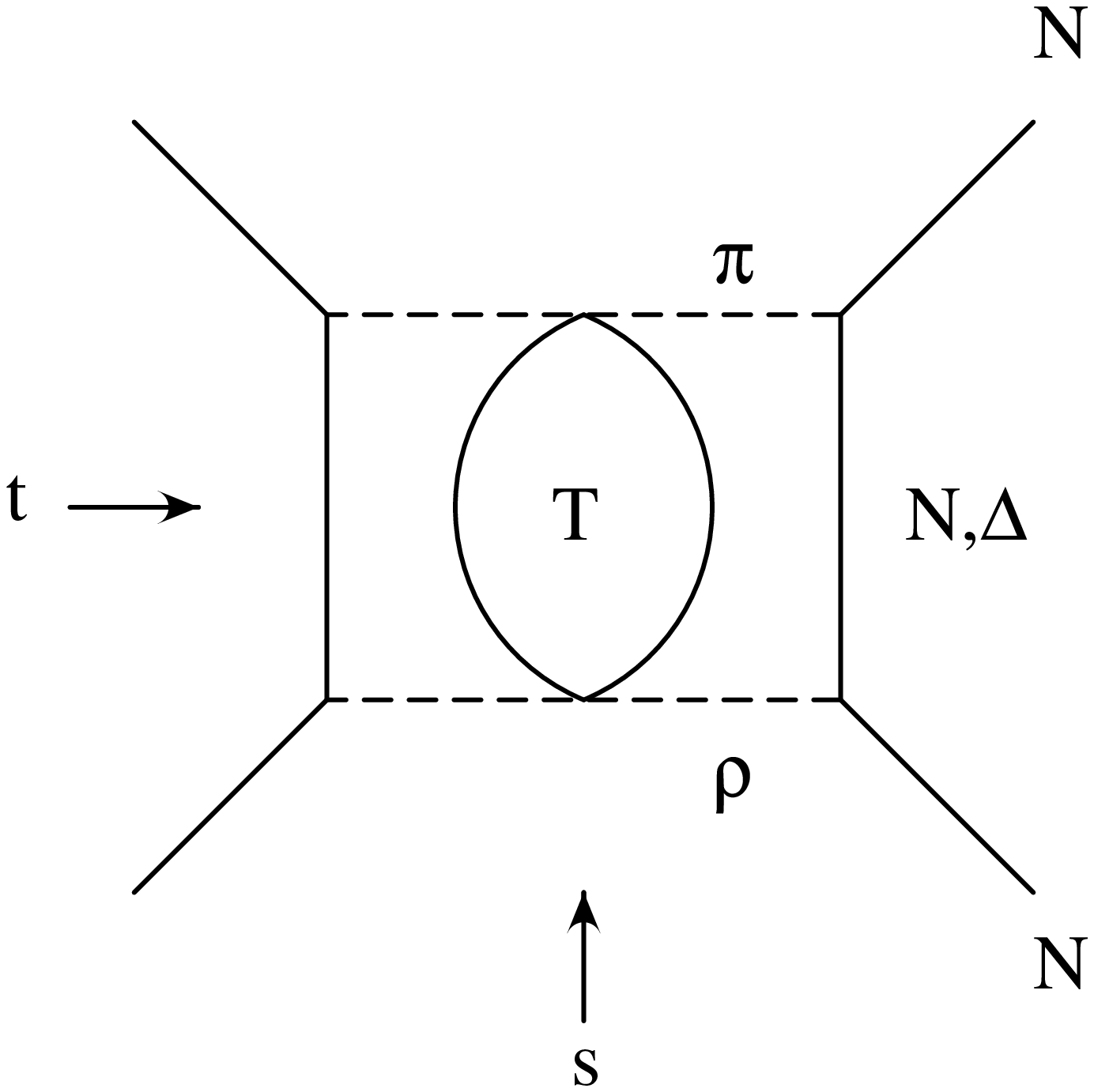}

\vskip 1cm

\centerline{\Large Figure 1}

\newpage

\vskip -2cm

\hskip 1.6in
\epsfysize=3in
\epsffile{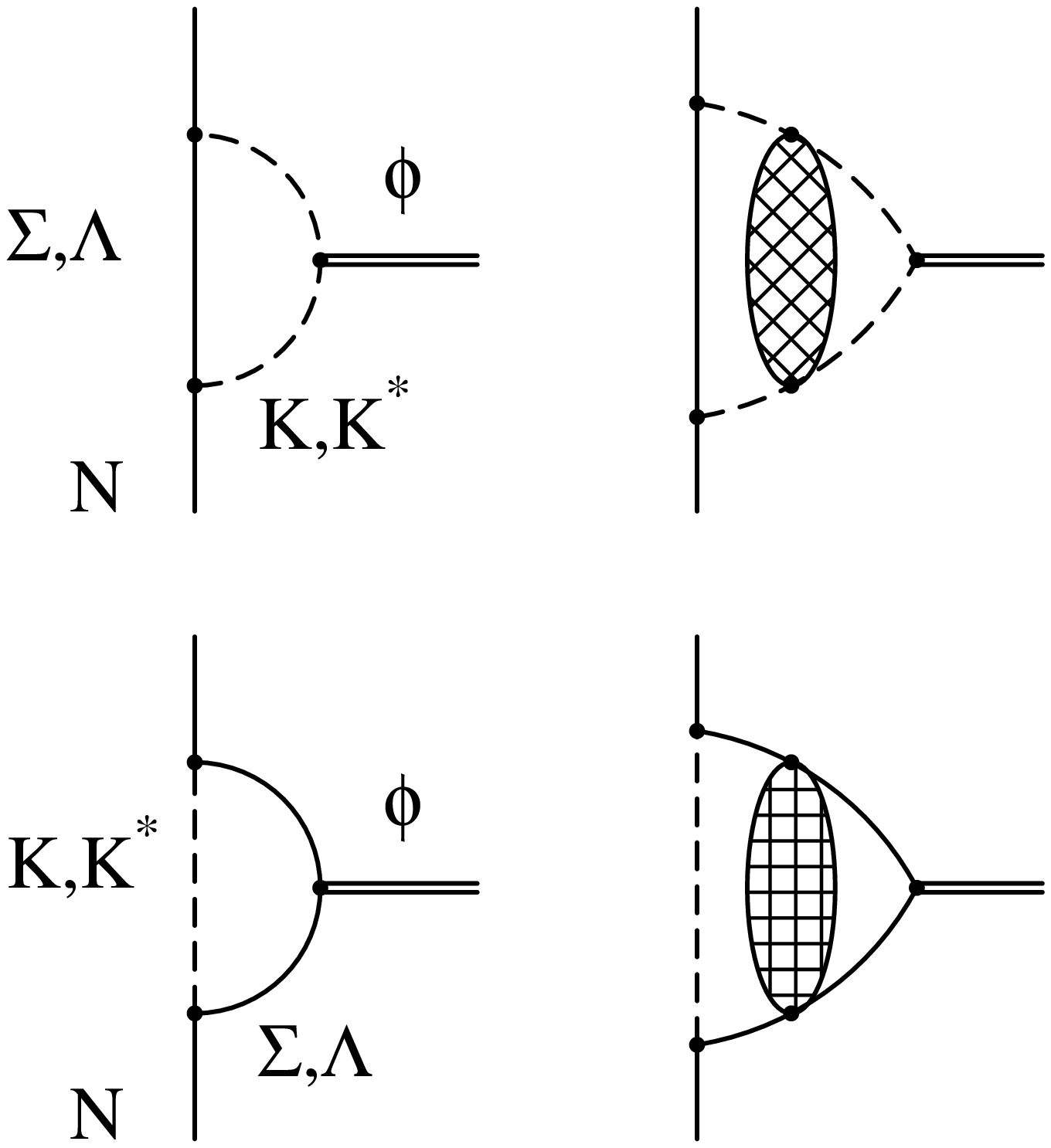}

\vskip 1cm

\centerline{\Large Figure 2}



\vskip 2cm

\hskip 1.in
\epsfxsize=3.7in
\epsffile{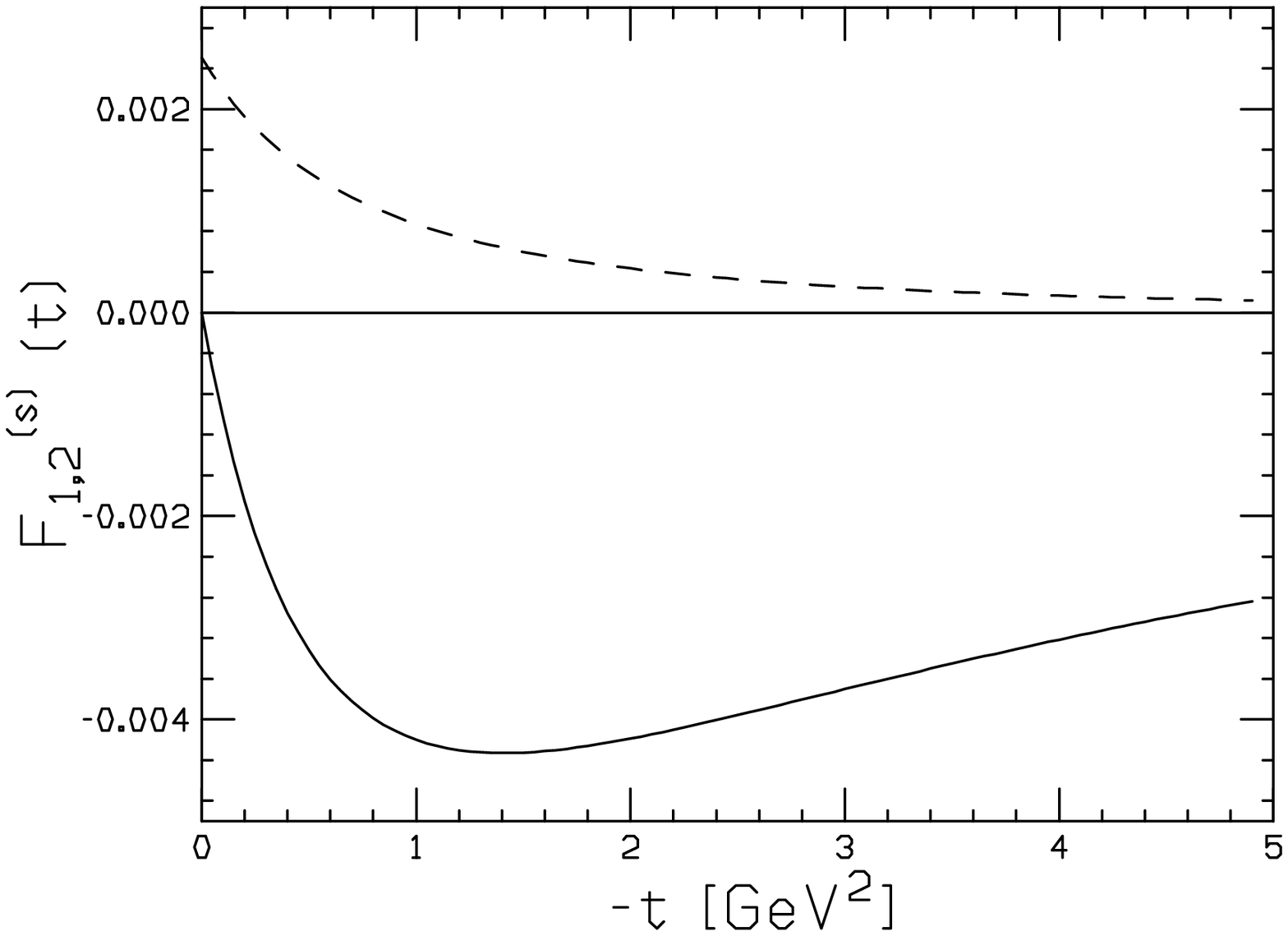} 

\vskip 2cm

\centerline{\Large Figure 3}


\begin{thebibliography}{99}
\bibitem{omff} Notice that in the dispersion--theoretical analyses of the
nucleons' electromagnetic form factors even larger values for $g_{\omega NN}$ 
are found, see \cite{hoeh76}\cite{mmd}.
\bibitem{nota} Our notation is identical to the one of \cite{mmd}, i.e.
the isoscalar ($I=0$) and isovector ($I=1$) form factors are given as 
half the sum and the difference of the proton and neutron ones, respectively.
\end{thebibliography}
\end{document}